\documentclass[fleqn,usenatbib]{mnras}

\usepackage{newtxtext,newtxmath}
\usepackage[T1]{fontenc}

\DeclareRobustCommand{\VAN}[3]{#2}
\let\VANthebibliography\thebibliography
\def\thebibliography{\DeclareRobustCommand{\VAN}[3]{##3}\VANthebibliography}

\usepackage{graphicx}	% Including figure files
\usepackage{amsmath}	% Advanced maths commands

\title[Cosmological non-ideal MHD waves]{Numerical tests of cosmological Alfvén waves with Ohmic diffusion}

\author[O. A. Karapiperis et al.]{
Orestis A. Karapiperis\thanks{E-mail: karapiperis@strw.leidenuniv.nl} \&
Matthieu Schaller
\\
Lorentz Institute for Theoretical Physics, Leiden University, PO Box 9506, NL-2300 RA Leiden, the Netherlands\\
Leiden Observatory, Leiden University, PO Box 9513, NL-2300 RA Leiden,
the Netherlands\\
}

\date{Accepted XXX. Received YYY; in original form ZZZ}

\pubyear{\the\year}

\begin{document}
\label{firstpage}
\pagerange{\pageref{firstpage}--\pageref{lastpage}}
\maketitle

\begin{abstract}
Physical problems with a solution that can be expressed analytically are scarce; this holds even more true for problems set in a cosmological context. Such solutions are, however, invaluable tools for making comparisons between theory, numerical experimentation and observations. In this work we present what to our knowledge is the first set of non-trivial closed-form expressions describing the behaviour of a system governed by the equations of non-ideal Magnetohydrodynamics (MHD), where the effects of Ohmic diffusion are considered, in a cosmologically expanding frame. We provide analytical solutions that describe the time evolution of linear perturbations to a homogeneous background in a radiation-dominated universe, yielding dissipative Alfvén waves. Although in our base framework solutions for any other cosmology of interest cannot be expressed in a closed form, they can still be obtained reliably through numerical integration of the coupled system of ordinary differential equations we provide. We compare our analytical solutions to numerical results obtained using our novel implementation of Smoothed Particle Magnetohydrodynamics (SPMHD) in the SWIFT astrophysical simulation code, to find good agreement between the two. We find the code to display good convergence behaviour, its predictions agreeing with theory to within $0.1\%$ for a modest number of resolution elements and at a negligible computational cost. We aim this work as a companion and supplement to the cosmological ideal MHD wave tests recently presented in the literature, and suggest that it be adopted as part of standard testing of code implementations of MHD.

\end{abstract}


%%%%%%%%%%%%%%%%%%%%%%%%%%%%%%%%%%%%%%%%%%%%%%%%%%

%%%%%%%%%%%%%%%%% BODY OF PAPER %%%%%%%%%%%%%%%%%%

\section{Introduction}
\label{sec:introduction}

Magnetic fields are ubiquitous in the universe, and their presence has been observationally inferred in objects ranging from planetary to galaxy cluster scales \citep{Beck2013}. They are of interest both in their own right, as their origin is still debated and presents a challenge for observers and theorists alike \citep{2021Galax...9..109V}, but also for their role in a wide variety of astrophysical phenomena (star formation, accretion onto compact objects, interstellar medium physics, circumgalactic medium physics, etc.). 

In a wide range of scientifically interesting instances, it is fair to assume that the astrophysical fluid under consideration behaves as a perfect conductor, in which case its evolution can accurately be described by the equations of ideal magnetohydrodynamics (MHD). These have been implemented in several astrophysical codes \citep{2006A&A...457..371F, 2008ApJS..178..137S, 2009MNRAS.398.1678D, 2011MNRAS.418.1392P, 2016MNRAS.455...51H, 2018PASA...35...31P}, which constitute the main theoretical tool used to study the evolution of magnetised gases. Such implementations have been subsequently used to successfully model systems ranging from gravitationally collapsing protostellar molecular cores (e.g. \citealt{2008A&A...477....9H}), to isolated Milky Way-like galaxies subject to supernova feedback \citep{2016MNRAS.455...51H}, to global populations of radio halos \citep{2018MNRAS.480.5113M}.

In certain settings however such as cold, dense environments, the assumption of infinite conductivity might be a poor approximation; one then needs to take into account non-ideal, first-order corrections to the equations of ideal MHD. The extension we will consider here is Ohmic diffusion, where collisions between charged particles diffuse the magnetic field away, affecting its topology, and inject thermal energy into the gas through Joule heating. Such processes are believed to be of importance in driving extreme space weather events in the Sun \citep{1983ApJ...264..642P}, permitting the settling of stable, rotationally supported discs around protostars \citep{2010ApJ...716.1541K}, or suppressing the launching of magnetically driven winds in star-forming environments \citep{2007ApJ...670.1198M}. Going beyond ideal MHD is relevant to galactic science as well, due to the tight link between magnetic diffusion and dynamo action (see~\citealt{2023ARA&A..61..561B} for a recent review) which is believed to take place in any turbulent astrophysical flow. Otherwise successful, self-consistent, contemporary galaxy formation models (e.g. \citealt{2014MNRAS.442.2304H, 2015MNRAS.446..521S, 2017MNRAS.467.4739K, 2019MNRAS.486.2827D} for pure hydrodynamical models, \citealt{2018MNRAS.473.4077P} who consider MHD in the ideal limit, or~\citealt{B_ss_2024} who use a solver that has been extended to include non-ideal MHD effects; see also~\citealt{Vazza_2017} for an extensive study of the impact of different astrophysical feedback mechanisms on large-scale magnetic fields) have still left long-standing open questions related to non-ideal MHD effects (such as the problem of catastrophic quenching of magnetic field growth, \citealt{1992ApJ...393..165V}) unexplored in the context of simulated galaxy populations. Smaller-scale simulations have nonetheless already shed light on the coupling between diffusive MHD, dynamos and astrophysical processes in galaxy formation, such as \cite{2021ApJ...910L..15G} who demonstrate that there exists a critical resistivity beyond which supernova-driven turbulence in the interstellar medium can excite a small-scale dynamo. Finally, works such as \cite{2011MNRAS.418.2234B} have investigated the impact of the inclusion of Ohmic resistivity in simulations of galaxy clusters, using it to reproduce magnetic field profiles derived from radio observations; the match occurs for a diffusion coefficient that is in line with theoretical estimates, based on the expected level of turbulent motion at their resolution scale.

The accuracy of the treatment of any physics newly included in a numerical solver is assessed through running a series of benchmark example problems with known analytical, or well-established numerical, solutions. Numerous works detail code implementations of non-ideal MHD and the standardised tests used to validate them. Focusing on approaches using the one-fluid approximation (see e.g.~\citealt{2008MNRAS.385.2269P}), whereby a single fluid comprised of both ions and neutrals is considered, examples include:~\cite{2012ApJS..201...24M} for an adaptive mesh refinement implementation,~\cite{2018MNRAS.476.2476M} and~\cite{2024MNRAS.527.1563Z} for moving mesh implementations, or~\cite{2013MNRAS.434.2593T} and~\cite{2014MNRAS.444.1104W} for the case of smoothed particle magnetohydrodynamics (SPMHD), which is the method that we will make use of in this work. Tests are usually presented in the method description literature in order of increasing complexity. One could typically encounter, in this order, simple wave propagation tests~\citep{2009ApJS..181..413C}, then shockwave tests~\citep{1980ApJ...241.1021D}, leading up to MHD instabilities~\citep{1963PhFl....6..459F} or gravitationally collapsing molecular clouds~\citep{2007ApJ...670.1198M}. Nonetheless, setups to assess whether an MHD solver correctly couples magnetic field physics to cosmology are still scarcely considered. MHD counterparts to the already challenging~\cite{1970A&A.....5...84Z} pancake and Santa Barbara cluster~\citep{1999ApJ...525..554F} tests are presented in~\cite{2008ApJS..174....1L} and~\cite{2011ApJS..195....5M} respectively. These problems are unquestionably valuable in stress testing, but as they are highly non-linear in nature and costly to run, they do not constitute a practical test-set suitable for evaluating code accuracy. It is with that in mind that~\cite{2022MNRAS.515.3492B} presented a novel set of analytical reference hydromagnetic wave solutions to the equations of cosmological MHD. They then proceeded to show how these can be reproduced (at very little computational cost) with the simulation code AREPO~\citep{2010MNRAS.401..791S} using the MHD implementation described in~\cite{2011MNRAS.418.1392P} and~\cite{2013MNRAS.432..176P} with the improvements described in~\cite{2016MNRAS.462.2603P}, and even more interestingly show how otherwise intractable implementation mistakes can be caught by running these test problems. We here set out to extend their work to non-ideal MHD, seeking counterpart analytical wave-like solutions for the case when Ohmic diffusion terms are included (noting that other dissipative effects on MHD modes have already been studied in the literature, such as viscous and heat conducting processes in~\cite{1998PhRvD..57.3264J} and~\cite{1998PhRvD..58h3502S}, or Navier-Stokes viscosity in~\cite{2022MNRAS.515.3492B}). We moreover test whether we can retrieve these using the novel implementation of cosmological SPMHD~(\citeauthor[][in prep.]{2025MNRAS}) in the simulation code SWIFT~\citep{2024MNRAS.530.2378S}. This exercise aims to evaluate whether we have encoded the equations of MHD in an expanding frame correctly, and whether their integration in time, inextricably linked to cosmology in SWIFT, is performed accurately. Concretely speaking, running this example with SWIFT serves for instance as an ideal testing bed for our choice of co-moving magnetic field variable evolved internally by the code; the actual physics considered being agnostic to this choice, retrieving expected results in the physical frame would serve as an indicator that our implementation choice has not introduced any unwanted effects.

This work is structured as follows: In Section~\ref{sec:mhd_theory} we first present the form of the equations of non-ideal MHD we use, we subsequently give the cosmological evolution we consider, and we finally show how we incorporate the latter into the former. In Section~\ref{sec:alfven_waves} we derive mathematical expressions describing the time evolution of the Fourier amplitude of standing cosmological dissipative Alfvén waves in a radiation-dominated universe. In Section~\ref{sec:alfven_waves_with_swift} we compare our analytical solutions to numerical results obtained with SWIFT, before discussing our findings and concluding in Section~\ref{sec:conclusion}.

\section{Non-ideal MHD in a co-moving frame}
\label{sec:mhd_theory}

We consider a fluid for which we relax the ideal MHD assumption of infinite electric conductivity $\sigma$, and allow for a non-zero Ohmic diffusion coefficient $\eta = 1 / \sigma$ that we take to be constant in both space and time. The time evolution of our system - in a Cartesian frame with spatial coordinates labelled as $\boldsymbol{r}$ - can then be taken to be governed by the mass continuity, momentum, induction and energy equations coupled to Newtonian gravity
\begin{equation}
    \frac{\mathrm{d} \rho}{\mathrm{d}t} 
    = - \rho \nabla \cdot \boldsymbol{v}
\label{eq:mass_continuity}    
\end{equation}
\begin{equation}
    \frac{\mathrm{d} \boldsymbol{v}}{\mathrm{d}t} 
    = - \frac{1}{\rho} \nabla_{\boldsymbol{r}} \cdot \boldsymbol{S}
    - \nabla_{\boldsymbol{r}} \Phi
\label{eq:momentum}
\end{equation}
\begin{equation}
    \frac{\mathrm{d}\boldsymbol{B}}{\mathrm{d}t} 
    = ( \boldsymbol{B} \cdot \nabla_{\boldsymbol{r}} ) \boldsymbol{v}
    - ( \nabla_{\boldsymbol{r}} \cdot \boldsymbol{v} ) \boldsymbol{B}
    + \eta \nabla^2 \boldsymbol{B}
\label{eq:induction}
\end{equation}
\begin{equation}
    \frac{\mathrm{d} u}{\mathrm{d}t}
    = - \frac{P}{\rho} \nabla_{\boldsymbol{r}} \cdot \boldsymbol{v}
    + \frac{\eta}{\rho} \boldsymbol{J}^2
\label{eq:energy}
\end{equation}
for $\mathrm{d}_t \equiv \partial_t + \boldsymbol{v} \cdot \nabla_{\boldsymbol{r}}$ the convective time derivative, $\rho$ the fluid density, $\boldsymbol{v} \equiv \mathrm{d}_t{\boldsymbol{r}}$ the fluid velocity, $u$ the specific internal energy, $P$ the thermal pressure, $\Phi$ the gravitational potential, and $\boldsymbol{B}$ the magnetic field. We have defined the Maxwell stress tensor
\begin{equation}
    S^{ij} \equiv \left( P + \frac{1}{2 \mu_0} B^2 \right) \delta^{ij} -
    \frac{1}{\mu_0} B^i B^j
\label{eq:maxwell_stress_tensor}
\end{equation}
for $\mu_0$ the permeability of free space, $\delta^{ij}$ the Kronecker delta, and the magnetic current
\begin{equation}
    \boldsymbol{J} \equiv \frac{1}{\mu_0} \nabla \times \boldsymbol{B}
\end{equation}
The fluid is moreover assumed to follow the ideal gas equation of state 
\begin{equation}
    P = \left( \gamma - 1 \right) \rho u
\label{eq:eos}
\end{equation}
with $\gamma$ the adiabatic index. The magnetic field satisfies the solenoidal constraint
\begin{equation}
    \nabla_{\boldsymbol{r}} \cdot \boldsymbol{B} = 0
\label{eq:solenoidal_constraint}
\end{equation}
and the gravitational potential follows from Poisson's equation
\begin{equation}
    \nabla_{\boldsymbol{r}}^2 \Phi = 4 \pi G \rho
\label{eq:poisson}
\end{equation}
for $G$ the gravitational constant.
We note that in the limit $\eta \rightarrow 0$, the system of equations above reduces to ideal MHD, as the diffusive terms in~(\ref{eq:induction}) and~(\ref{eq:energy}) vanish.

To derive the time evolution of our system in an expanding frame, we start by considering co-moving spatial coordinates labelled as $\boldsymbol{x}$, defined through $\boldsymbol{r} = a(t) \boldsymbol{x}$. Here $a(t)$ is the cosmological scale factor whose time evolution we take to be governed by the Friedmann equation
\begin{equation}
    H(a) = H_0
    \left[
    \Omega_m a^{-3} + \Omega_r a^{-4} + \Omega_\Lambda
    \right]^{1/2}
\label{eq:friedmann}
\end{equation}
where $H \equiv \dot{a} / a$ is the Hubble parameter, $H_0 = 100 \: h^{-1} \: \mathrm{km} \: \mathrm{s}^{-1} \: \mathrm{Mpc}^{-1}$ is its present day value and $h$ its reduced counterpart. $\Omega_m$, $\Omega_r$, and $\Omega_\Lambda$ are respectively the present-day dimensionless matter, radiation, and dark energy densities. We further define a co-moving counterpart $Q_c$ to any physical quantity $Q$ as $Q_c = a^n Q$ for some $n \in \mathbb{R}$ ; we make the choices 
\begin{equation}
    \rho_c = a^3 \rho \text{,}
    \quad P_c = a^{3 \gamma} P \text{,}
    \quad u_c = a^{3(\gamma-1)} u \text{,}  
    \quad \boldsymbol{B}_c = a^{3 \gamma / 2} \boldsymbol{B}
\label{eq:comoving_variables}
\end{equation}
which ensure that thermodynamic relations are the same in the expanding and co-moving frames\footnote{Note that there is no single convention for defining a co-moving magnetic field; our choice nonetheless differs from that most commonly employed~\citep[e.g.][]{2011MNRAS.418.1392P}, namely $\boldsymbol{B}_c = a^2 \boldsymbol{B}$.}. We further introduce the velocity variable $\boldsymbol{w} = a^2 \dot{\boldsymbol{x}}$ (noting that it has no clear physical interpretation as it does not match the peculiar velocity $\boldsymbol{u} \equiv a \dot{\boldsymbol{x}}$ or any other commonly employed physical velocity) and plug~(\ref{eq:comoving_variables}) into~(\ref{eq:mass_continuity})-(\ref{eq:energy}) to get the equations of non-ideal MHD in a co-moving frame
\begin{equation}
    \frac{\mathrm{d} \rho_c}{\mathrm{d}t} 
    = - \frac{1}{a^2} \rho_c \nabla_{\boldsymbol{x}} \cdot \boldsymbol{w}
\label{eq:comoving_mass_continuity}
\end{equation}
\begin{equation}
    \frac{\mathrm{d} \boldsymbol{w}}{\mathrm{d}t} 
    = - \frac{1}{a^{3(\gamma-1)}} 
    \frac{1}{\rho}_c \nabla_{\boldsymbol{x}} \cdot \boldsymbol{S}_c
    - \frac{1}{a} \nabla_{\boldsymbol{x}} \Phi_c
\label{eq:comoving_momentum}
\end{equation}
\begin{equation}
    \frac{\mathrm{d}\boldsymbol{B}_c}{\mathrm{d}t} 
    = \frac{1}{a^2} \bigg\{ 
    ( \boldsymbol{B}_c \cdot \nabla_{\boldsymbol{x}} ) \boldsymbol{w}
    - ( \nabla_{\boldsymbol{x}} \cdot \boldsymbol{w} ) \boldsymbol{B}_c
    + \eta \nabla^2 \boldsymbol{B}_c
    \bigg\}
    + \Gamma H \boldsymbol{B}_c
\label{eq:comoving_induction}
\end{equation}
\begin{equation}
    \frac{\mathrm{d} u_c}{\mathrm{d}t}
    = \frac{1}{a^2} \bigg\{
    - \frac{P_c}{\rho_c} \nabla_{\boldsymbol{x}} \cdot \boldsymbol{w}
    + \frac{\eta}{\rho_c} \boldsymbol{J}_c^2
    \bigg\}
\label{eq:comoving_energy}
\end{equation}
where $\boldsymbol{J}_c = \frac{1}{\mu_0} \nabla_{\boldsymbol{x}} \times \boldsymbol{B}_c$ and for convenience we have defined $\Gamma \equiv 3 \gamma / 2 - 2$ and the co-moving potential $\Phi_c = a \Phi + \frac{1}{2}a^2 \ddot{a} \boldsymbol{x}^2$. We note that our choice of co-moving variables introduces a source term in the induction equation; this will prove to be unimportant in what follows.

\section{Cosmological Alfvén waves with Ohmic diffusion}
\label{sec:alfven_waves}

We will now proceed with the standard perturbation theory exercise of deriving the time evolution equations for small deviations of physical variables of interest from a uniform, static background. We closely follow the procedure outlined by~\cite{2022MNRAS.515.3492B}, using their notation, for ease of comparison and with the aim in mind of their analytical solutions and ours being considered jointly for testing code implementations of cosmological MHD.

\subsection{A background to perturb on}
\label{sec:background}

The ground state we start from is static, i.e. $\boldsymbol{w} = \boldsymbol{0}$, and has spatially uniform $\rho_c$, $P_c$ and $\boldsymbol{B}_c$. By inspection of equations~(\ref{eq:momentum})-(\ref{eq:energy}) it becomes evident that $\rho_c$, $P_c$ and $u_c$ remain constant in time as all spatial gradients vanish, while the background magnetic field evolves monotonically as
\begin{equation}
    \boldsymbol{B}_c (a) = a^\Gamma  \boldsymbol{B}_{c,z=0}
\label{eq:background_B_evolution}
\end{equation}
where we use redshift $z \equiv 1/a - 1$ to denote the present day, when (by definition) $z=0$ and the background magnetic field reaches the value $\boldsymbol{B}_{c,z=0}$. This is a slight departure from~\cite{2022MNRAS.515.3492B}, where $\boldsymbol{B}_c$ remains constant in time and $P_c$ decays for non-isothermal gases, owing to our different choices of co-moving variables and (nonetheless physically equivalent) ways of tracking gas thermodynamics.

\subsection{Characteristic speeds}
\label{sec:characteristic_speeds}

Again following \cite{2022MNRAS.515.3492B}, but adapting as appropriate to our formulation of the physics outlined in Section~\ref{sec:mhd_theory} and extending to non-zero $\eta$, we define a set of characteristic speeds that naturally arise in non-ideal MHD and that will considerably help compactify expressions later on. We define the co-moving Alfvén speed and characteristic velocity for magnetic diffusion $\mathcal{V}_A$ and $\mathcal{V}_\eta$ as
\begin{equation}
    \mathcal{V}_A \equiv \frac{|\boldsymbol{B}_{c,z=0}|}{\sqrt{\mu_0 \rho_c}}
    \quad \text{and} \quad
    \mathcal{V}_\eta \equiv k \eta
\label{eq:sound_and_alfven_speeds}
\end{equation}
respectively, for $k$ the co-moving wave vector of the wave-like solution we will derive in the following subsection. We additionally define dimensionless counterparts to these characteristic speeds as
\begin{equation}
    \Omega_A \equiv \frac{k \mathcal{V}_A}{H_0} , \quad
    \Omega_\eta \equiv \frac{k \mathcal{V}_\eta}{H_0}
\label{eq:dimensionless_speeds}
\end{equation}

\subsection{Cosmological Alfvén waves with Ohmic diffusion}
\label{sec:analytic_solutions}

We consider perturbations to the background presented in Section~\ref{sec:background} that will yield linearly polarised waves propagating in the $x$ direction. Without loss of generality, the background magnetic field is taken to point in the $x$ direction, i.e. $\boldsymbol{B}_c = B_c \hat{\boldsymbol{e}}_x$. We constrain the perturbations to be perpendicular to the propagation direction. For the magnetic and velocity fields, respectively, we let these take the form
\begin{equation}
    \delta \boldsymbol{B}_c = \delta B_c e^{ikx} \hat{\boldsymbol{e}}_y
    \quad \text{and} \quad
    \delta \boldsymbol{w} = \delta w e^{ikx} \hat{\boldsymbol{e}}_y
\label{eq:alfven_perturbations}
\end{equation}
where we allow the complex amplitudes $\delta B_c$ and $\delta w$ to vary in time. We also define the peculiar velocity perturbation $\delta \boldsymbol{u} = \frac{1}{a} \delta \boldsymbol{w}$, which we will later use to express results in terms of a physically meaningful quantity. The real space solution can then simply be retrieved by taking the real part of expression~(\ref{eq:alfven_perturbations}), e.g. $\boldsymbol{B}_c (x, t) = \Re(\boldsymbol{B}_c + \delta \boldsymbol{B}_c)$. We note that our choices do indeed yield a solenoidal magnetic field ($\nabla_{\boldsymbol{x}} \cdot \boldsymbol{B}_c=0$) and moreover lead to an incompressible flow ($\nabla_{\boldsymbol{x}} \cdot \boldsymbol{w} = 0$). The latter has as a consequence that, to linear order in the perturbations, the right-hand sides of equations~(\ref{eq:comoving_mass_continuity}) and~(\ref{eq:comoving_energy}) are zero and thus $\rho_c$, $P_c$ and $u_c$ can be taken to remain constant.

Given~(\ref{eq:alfven_perturbations}), equations~(\ref{eq:comoving_momentum}) and~(\ref{eq:comoving_induction}) reduce to
\begin{equation}
    \frac{\mathrm{d} \delta w}{\mathrm{d} t} = 
    \frac{1}{a}
    i k \frac{B_c^2}{\mu_0 \rho} \frac{\delta B_c}{B_c}
\label{eq:perturbed_momentum}
\end{equation}
and
\begin{equation}
    \frac{\mathrm{d}}{\mathrm{d} t}
    \left( \frac{\delta B_c}{B_c} \right) = \frac{1}{a^2} 
    \bigg\{
     i k \delta w - k^2 \eta \frac{\delta B_c}{B_c}
    \bigg\}
\label{eq:perturbed_induction}
\end{equation}
respectively.
To make further progress and arrive at analytically tractable solutions, we need to specify a cosmology, i.e. close off the system of coupled differential equations above by specifying a choice for the parameters $(\Omega_m, \Omega_r, \Omega_\Lambda)$ that would dictate the exact form of $\dot{a}$; \cite{2022MNRAS.515.3492B} choose to proceed with an Einstein-de Sitter (EdS) universe (i.e. $\Omega_m = 1$, with with $\Omega_r$ and $\Omega_\Lambda$ set to zero set to zero). Making this choice in our case, due to the additional terms brought about by the inclusion of Ohmic diffusion, leads to a solution that is fairly challenging to manipulate. We nonetheless present it in Appendix~\ref{app:alfven_waves_eds} and show that it reduces to the expression derived in \cite{2022MNRAS.515.3492B} in the limit $\eta \rightarrow 0$. This solution does simplify considerably if we relax the assumption that $\eta$ be scale factor independent. Even though this would only translate to a trivial modification of our base solver, it does not constitute a feature we wish to include in our general code release; we nevertheless implement it and present relevant results in Appendix~\ref{app:alfven_waves_eds_bis}, but note that they are not reproducible using an `out-of-the-box' clone of SWIFT.

For our current purposes, we choose to rather consider a radiation dominated universe (i.e. $\Omega_r = 1$, with $\Omega_m$ and $\Omega_\Lambda$ set to zero), for which $\dot{a} = H_0 / a$. We emphasise that this choice is not based on any physical argument and is most probably implausible, but nonetheless serves the primary purpose of the present exercise which is to derive analytically tractable solutions that cosmological MHD codes can be tested against. Taking the time variable to be the scale factor itself and using our definitions~(\ref{eq:dimensionless_speeds}), equations~(\ref{eq:perturbed_momentum}) and~(\ref{eq:perturbed_induction}) transform to
\begin{equation}
    \frac{\mathrm{d} \delta w}{\mathrm{d} a} = 
    \frac{i H_0 \Omega_A^2}{k} \frac{\delta B_c}{B_c}
\label{eq:pertubed_momentum_radiation_domination}
\end{equation}
and
\begin{equation}
    \frac{\mathrm{d}}{\mathrm{d} a}
    \left( \frac{\delta B_c}{B_c} \right) = 
    \frac{1}{a} \bigg\{ \frac{i k }{H_0} \delta w
    - \Omega_\eta \frac{\delta B_c}{B_c}
    \bigg\}
\label{eq:pertubed_induction_radiation_domination}
\end{equation}
respectively. These can then be combined to yield
\begin{equation}
    \frac{\mathrm{d}^2}{\mathrm{d} a^2}
    \left( \frac{\delta B_c}{B_c} \right)
    + \frac{1 + \Omega_\eta}{a} \frac{\mathrm{d}}{\mathrm{d} a}
    \left( \frac{\delta B_c}{B_c} \right)
    + \frac{\Omega_A^2}{a} \frac{\delta B_c}{B_c} = 0
\label{eq:master_ode_a}
\end{equation}
which resembles a damped oscillator equation, with non-constant coefficients. The restoring force scales like $\Omega_A$ and decays with the expansion of the universe (for all cosmologies we are interested in $a(t)$ is a monotonically increasing function of time). Moreover, the equation features a cosmological drag term that decreases with increasing $a(t)$, and is enhanced by the inclusion of $\Omega_\eta \neq 0$. After the coordinate transformation $\kappa = 2 \Omega_A \sqrt{a}$, equation~(\ref{eq:master_ode_a}) takes the form of the transformed Bessel's equation
\begin{equation}
    \frac{\mathrm{d}^2}{\mathrm{d} \kappa^2}
    \left( \frac{\delta B_c}{B_c} \right)
    + \frac{1 + 2 \Omega_\eta}{\kappa} \frac{\mathrm{d}}{\mathrm{d} \kappa}
    \left( \frac{\delta B_c}{B_c} \right)
    + \frac{\delta B_c}{B_c} = 0
\label{eq:master_ode_kappa}
\end{equation}
which has as a solution\footnote{This solution is comparable from a mathematical standpoint to the magnetosonic wave solution presented in \cite{2022MNRAS.515.3492B}; see their Appendix C3 or refer to e.g. \cite{1958ibf..book.....B} for details on how such equations can be solved.}
\begin{equation}
    \frac{\delta B_c}{B_c} (a) = 
    c_1 \mathcal{F} (a ; \Omega_A, \Omega_\eta) +
    c_2 \mathcal{G} (a ; \Omega_A, \Omega_\eta)
\label{eq:alfven_wave_B_solution}
\end{equation}
where $c_1$ and $c_2$ are integration constants, and where we have defined the functions $\mathcal{F}$ and $\mathcal{G}$ as
\begin{equation}
    \mathcal{F} (a ; \Omega_A, \Omega_\eta) = 
    a^{- \Omega_\eta / 2} J_{\Omega_\eta} (2 \Omega_A \sqrt{a})
\label{eq:definition_F}
\end{equation}
and
\begin{equation}
    \mathcal{G} (a ; \Omega_A, \Omega_\eta) = 
    a^{- \Omega_\eta / 2} Y_{\Omega_\eta} (2 \Omega_A \sqrt{a})
\label{eq:definition_G}
\end{equation}
for Bessel functions of the first and second kind of order $\alpha$, $J_\alpha$ and $Y_\alpha$, respectively. We note that the ratio $\delta B / B$ takes the same value in both the expanding and non co-moving frames meaning it has a direct physical interpretation, as should all output variables used in code evaluation and comparison projects. In an attempt to give some intuitive insight into~(\ref{eq:alfven_wave_B_solution}), we note that the solution obtained exhibits oscillatory behaviour, with decreasing frequency for larger $\Omega_\eta$\footnote{In the sense that, as we incrementally increase the order of a Bessel function, its $s$-th zero occurs for a larger argument; see e.g. \citealt{1972hmfw.book.....A}.}. This is then further modulated by an exponentially decaying envelope that itself depends on $\Omega_\eta$. Both features are also present, qualitatively, in resistive magnetised flows in a non-cosmological setting (see e.g. \citealt{2007MNRAS.381..319L}). 

Given the above, we can also retrieve an expression for $\delta w$ (or equivalently for $\delta u$) by combining~(\ref{eq:perturbed_induction}) and~(\ref{eq:alfven_wave_B_solution}) and using our definitions~(\ref{eq:dimensionless_speeds}), for a radiation dominated universe
\begin{align}
\begin{split}
    \frac{\delta w}{\mathcal{V}_A} (a)
    = & - \frac{i c_1}{\Omega_A} \bigg\{ 
    a \mathcal{F}' (a) + \Omega_\eta \mathcal{F} (a)
    \bigg\} \\
    & - \frac{i c_2}{\Omega_A} \bigg\{
    a \mathcal{G}' (a) + \Omega_\eta \mathcal{G} (a)
    \bigg\}
\label{eq:alfven_wave_w_solution}
\end{split}
\end{align}
where we use primes to denote differentiation with respect to $a$ and leave out the explicit dependence of $\mathcal{F}$ and $\mathcal{G}$ on $\Omega_A$ and $\Omega_\eta$ for clarity.

\begin{figure*}
 \includegraphics[]{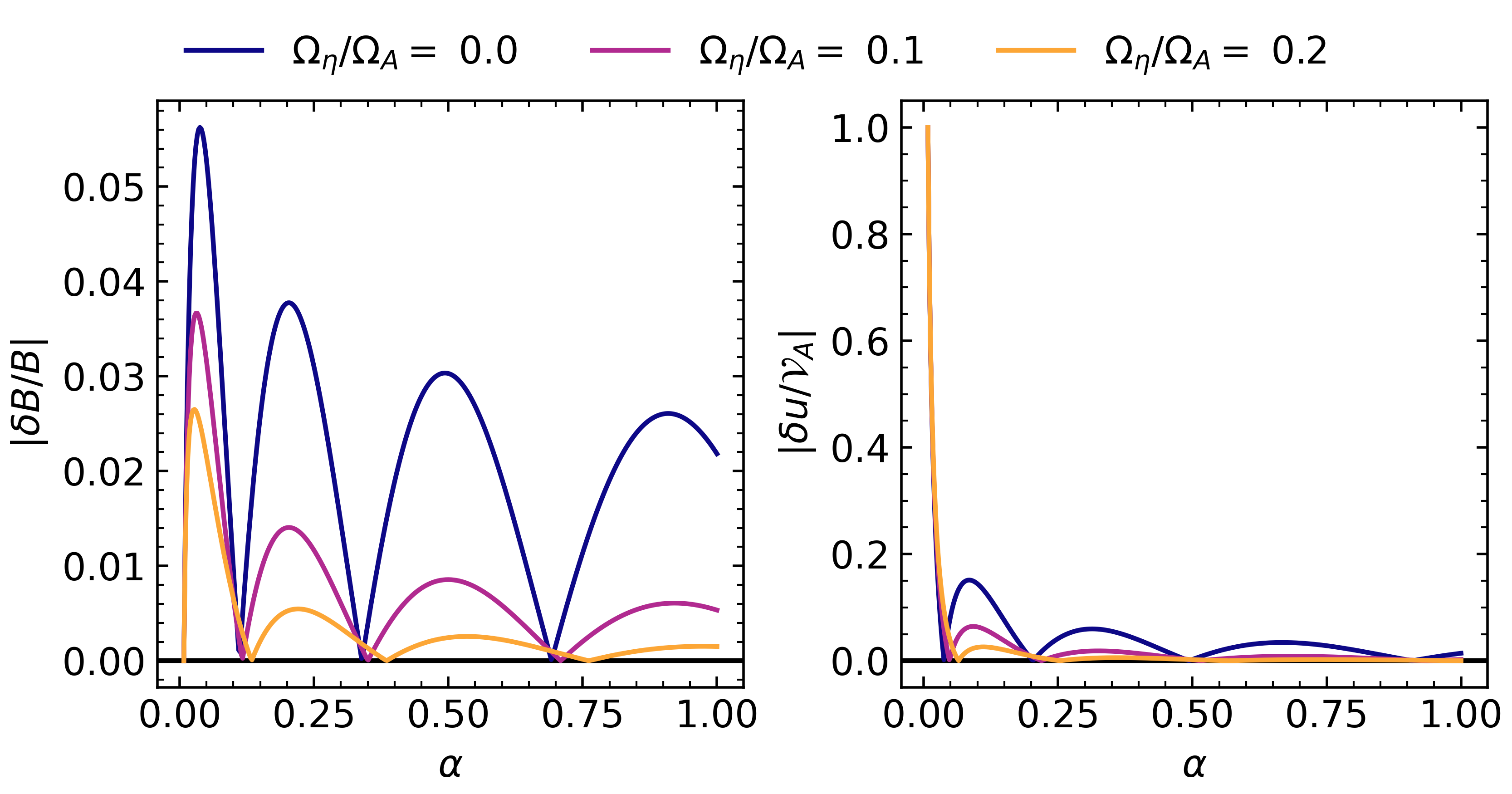}
 \caption{Time evolution of the Fourier amplitude of the magnetic field and velocity perturbations, $\delta B$ and $\delta u$ (equations~(\ref{eq:alfven_wave_B_solution}) and~(\ref{eq:alfven_wave_w_solution}) respectively), for different diffusion strengths (as expressed through the dimensionless diffusive speed $\Omega_\eta$). $\delta B$ is scaled to the background co-moving magnetic field $B_c$, while $\delta u$ is scaled to the background co-moving present-day Alfvén speed $\mathcal{V}_A$. We clearly observe an increasingly stronger damping of the wave for larger $\Omega_\eta$, while the frequency of oscillations decreases.}
 \label{fig:analytic_solution}
\end{figure*}

We now only have to specify the integration constants which, like \cite{2022MNRAS.515.3492B}, we obtain through solving~(\ref{eq:alfven_wave_B_solution}) and~(\ref{eq:alfven_wave_w_solution}) evaluated at our starting scale factor $a_i$ as set of linear equations for the two unknowns $c_1$ and $c_2$, in terms of the amplitude of the relative magnetic and velocity field perturbations at $a_i$
\begin{equation}
    A_B \equiv \frac{\delta B_c}{B_c} (a_i), \quad
    A_u \equiv \frac{\delta u}{\mathcal{V}_A} (a_i) .
\label{eq:initial_alfven_amplitudes}
\end{equation}
This exercise results in
\begin{align}
\begin{split}
    c_1 & = \pi a_i^{\Omega_\eta + 1} \bigg\{ A_B 
    \mathcal{G}'(a_i)
    - 
    \left( i \Omega_A A_w - \Omega_\eta A_B
    \right) 
    \frac{\mathcal{G}(a_i) }{a_i} \bigg\} \\
    c_2 & =  - \pi a_i^{\Omega_\eta + 1} \bigg\{ A_B 
    \mathcal{F}'(a_i)
    -
    \left( i \Omega_A A_w - \Omega_\eta A_B
    \right) 
    \frac{\mathcal{F}(a_i)}{a_i} \bigg\} .
\label{eq:alfven_integration_constants}
\end{split}
\end{align}
We note that these expressions simplify considerably for $A_B = 0$, i.e. for the case of an initially homogeneous magnetic field. Fig.~\ref{fig:analytic_solution} shows the time evolution of the Fourier amplitude of the magnetic field and velocity perturbations for such a choice of initial conditions, and for a range of ratios $\Omega_\eta / \Omega_A$ where $\Omega_A$ is kept fixed; we clearly observe both the damping of the wave and the shift of its zeroes to larger-scale factors as the strength of Ohmic dissipation is increased.

\section{Simulating diffusive cosmological Alfvén waves with SWIFT}
\label{sec:alfven_waves_with_swift}

We present here a comparison between the analytical solutions derived in Section~\ref{sec:alfven_waves} and the results of numerical experiments performed with the simulation code SWIFT, after briefly introducing our software and numerical method of choice, and the set of model hyperparameters we ran with. For a full description of the methods employed, we refer the reader to the works cited in the two following subsections.

\subsection{The SWIFT simulation code}

SWIFT~\citep{2024MNRAS.530.2378S} is a fully open-source and highly parallel software package that jointly solves the equations of magnetohydrodynamics and gravity in a cosmological context; it additionally comes with a large selection of effective models (subgrid models), which account for the impact of astrophysical phenomena that occur below the resolution scale and are relevant for galaxy formation. The code makes use of modern computational approaches and algorithms such as task-based parallelism, fully dynamic and asynchronous communication and graph-based domain decomposition to make optimal use of modern hybrid shared / distributed-memory computer architectures.

\subsection{SPMHD in SWIFT}

Gas physics is accounted for in SWIFT using Smoothed Particle Hydrodynamics (SPH), a Lagrangian particle-based method that discretises the fluid under consideration on mass and offers good conservation properties, stability, and naturally adaptive resolution at a low computational cost. SWIFT includes implementations of several flavours of SPH; here we make use of the state-of-the-art SPHENIX scheme~\citep{2022MNRAS.511.2367B}, which incorporates mitigation strategies that address most shortcomings of SPH, and which was specifically designed with cosmology and galaxy formation applications in mind.

We further make use of the SPMHD solver described in~\citeauthor[][in prep.]{2025MNRAS}, to account for magnetic field physics. This solver includes the latest suggested versions of corrective measures which have allowed for stable and reliable implementations of SPMHD, making the method applicable to a wide range of astrophysical study cases (for a recent review see~\citealt{2023FrASS..1088219T}):
\begin{enumerate}
    \item A formulation of the equations of motion derived from an action minimisation principle~\citep{2004MNRAS.348..139P}, which conserves energy and linear momentum.
    \item A correction to the acceleration calculation, cancelling spurious magnetic forces arising numerically and making the method formally equivalent to the eight-wave cleaning approach of~\cite{1999JCoPh.154..284P}.
    \item The `artificial resistivity' prescription described in~\cite{2018PASA...35...31P} to smooth magnetic field discontinuities over the resolution scale.
    \item The constrained hyperbolic divergence cleaning scheme of~\cite{2016JCoPh.322..326T} to control magnetic divergence errors.
\end{enumerate}
Hyperparameter values for the models listed above were chosen in SWIFT based on an extensive and exhaustive numerical experimentation campaign using targeted, idealised test problems. Details can be found in~\citeauthor[][in prep]{2025MNRAS}.

Our SPMHD calculations were performed using a 4th order B-spline \citep[as in][]{2022MNRAS.511.2367B} as the smoothing kernel with a resolution parameter set to $\eta=1.2$, corresponding to $\sim 57$ weighted particle neighbours being considered in the interpolation sums. Neighbour finding being the most computationally demanding task in SPH calculations, it is worth noting that other works testing implementations of SPMHD~\citep{2018PASA...35...31P, 2020A&A...638A.140W} make use of considerably larger kernels; we find this not to be necessary here and reassert that the aim of the present exercise is to devise inexpensive tests that streamline and accelerate evaluating code performance. 

In our runs, all solver hyperparameters were set to their default values as given in~\cite{2022MNRAS.511.2367B} and~\citeauthor[][in prep.]{2025MNRAS}, and time-step sizes used in the time integration were calculated through the Courant–Friedrichs–Lewy condition~\citep{1928MatAn.100...32C}, computed using a prefactor of $C_{\mathrm{CFL}}=0.1$ and the fast magnetosonic wave speed as the maximal information propagation velocity. We moreover considered, as is common practice, a diffusion-based time-step criterion for non-ideal runs; this did not prove prohibitive, so we did not have to resort to a super-timestepping algorithm (as in e.g.~\citealt{2024MNRAS.527.1563Z})

\subsection{Diffusive cosmological Alfvén waves in a radiation-dominated universe}

\subsubsection{Initialisation}
\label{sec:initialisation}

We now proceed to a description of the procedure we followed to set up standing, linearly polarised, cosmological Alfvén waves with SWIFT. The derivation presented in Section~\ref{sec:alfven_waves} is agnostic to the exact scale of most of the attributes of the gas ($\rho, P, u$, etc), and does not impose any restrictions on the amplitude of the perturbations in $u$ and $\boldsymbol{B}$ as long as these remain within the regime of validity of linear perturbation theory. We, therefore, choose to run with physical parameters either matched to, or of the same order of magnitude as those used in the cosmological perturbation growth study of  ~\cite{1970A&A.....5...84Z}. This set of parameters has both the advantage of being physically plausible and, moreover, belongs to a regime extensively explored by the community for cosmological setups.

We initialise $64 \times 16 \times 16$ particles on a uniform, three-dimensional Cartesian grid in a periodic simulation box with a major axis of length $L = 64 \: \mathrm{Mpc} \: h^{-1}$, setting $h=1$ for simplicity. Running in 3D is a notable departure from the approach of~\cite{2022MNRAS.515.3492B}, who show results from 1D and convergence studies from 2D runs. Much like~\cite{2009MNRAS.398.1678D},~\cite{2018PASA...35...31P} or~\cite{2022MNRAS.511.2367B}, we deem it useful to test the code in a configuration close to the one meant for production runs, even for tests of lower dimensionality. 

We set $\gamma = 5/3$ and give the gas an initial mass density of $\rho_i = 3 H_0^2 / 8 \pi G$, an initial temperature of $T_i = 100 \: \mathrm{K}$ and choose the initial, uniform background magnetic field so that $\Omega_A = 2 \pi$. The amplitude of velocity perturbations introduced on top of the fluid's static ground state is fixed so that $A_u = 1.0$. We choose to consider standing waves of a single period with wave-number $k = 2 \pi / L$, yielding $\delta u (a_i) = A_u \mathcal{V}_A \mathrm{cos} (k x)$, with $x$ lying along the major axis of the simulation box. We then let the system evolve from $z=127$ to $z=0$, having configured the code to consider a cosmology corresponding to a radiation-dominated universe. We repeat the experiment three times, for three different magnetic diffusion strengths $\eta$, as expressed through the ratio $\Omega_\eta / \Omega_A$ ($\Omega_A $ is kept fixed for all runs). We run with $\Omega_\eta / \Omega_A = {0, \, 0.1, \, 0.2}$ which constitutes a set of values through which the impact of magnetic diffusion on the time evolution of cosmological Alfvén waves is made evident\footnote{Both the MHD module and diffusive cosmological Alfvén wave example will soon be made publicly available, in a forthcoming SWIFT code version release.}.

\subsubsection{Results}

\begin{figure*}
 \includegraphics[]{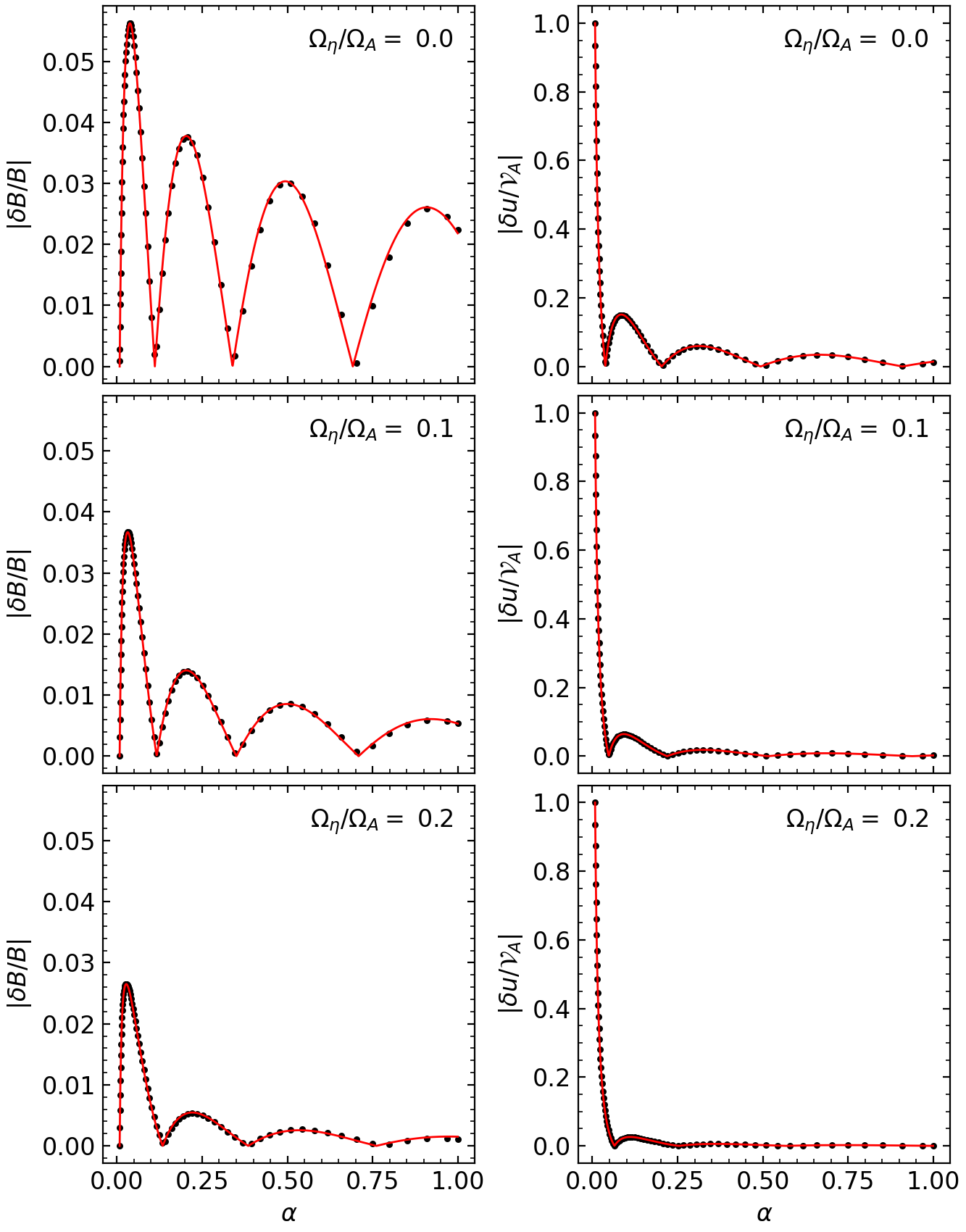}
 \caption{Analytical solutions (red line) plotted against numerical results obtained with SWIFT (black dots) for the time evolution in a radiation-dominated universe of the Fourier amplitude of magnetic (left column) and velocity (right column) linear perturbations corresponding to a standing, linearly polarised, cosmological Alfvén wave initialised as described in Section~\ref{sec:initialisation}. Individual rows correspond to different strengths of the Ohmic diffusion coefficient $\eta$ as expressed through the ratio $\Omega_\eta / \Omega_A$. The numerical solution agrees well with the analytically derived result, showing an increasingly strong damping (reduction of the wave amplitude) and dispersive (shift of zeroes to the right) effect as $\eta$ is increased.}
 \label{fig:results}
\end{figure*}

In Figure~\ref{fig:results} we present a comparison of the analytical solution presented in Section~\ref{sec:alfven_waves} to the numerical results obtained with SWIFT for a standing, linearly polarised, cosmological Alfvén wave initialised as described in Section~\ref{sec:initialisation}. We show the evolution, as a function of scale factor $a$ and for different Ohmic diffusion strengths $\eta$, of the Fourier amplitude of magnetic and velocity linear perturbations $\delta B_c$ and $\delta u$, scaled to the background magnetic field $B_c$ and present-day Alfvén velocity $\mathcal{V}_A$ respectively. The code results were generated by reading the calculated magnetic and velocity fields along $\boldsymbol{e}_y$ (the direction along which the linear perturbations lie) from SWIFT outputs, and extracting the amplitudes of the observed sine and cosine profiles respectively. 
 
The solution takes the form of a dispersive oscillation, modulated by a decaying envelope, for both $\delta B_c$ and $\delta u$. We clearly observe the trend that was anticipated given the form of the analytical solution: increasing the Ohmic diffusion strength leads to increasingly prominent damping and dispersive effects. The former leads to a reduction of the wave amplitude at all $a$ for larger $\eta$. The latter is made evident through a shift to larger $a$ of the zeroes of the wave amplitudes. 

\begin{figure*}
 \includegraphics[]{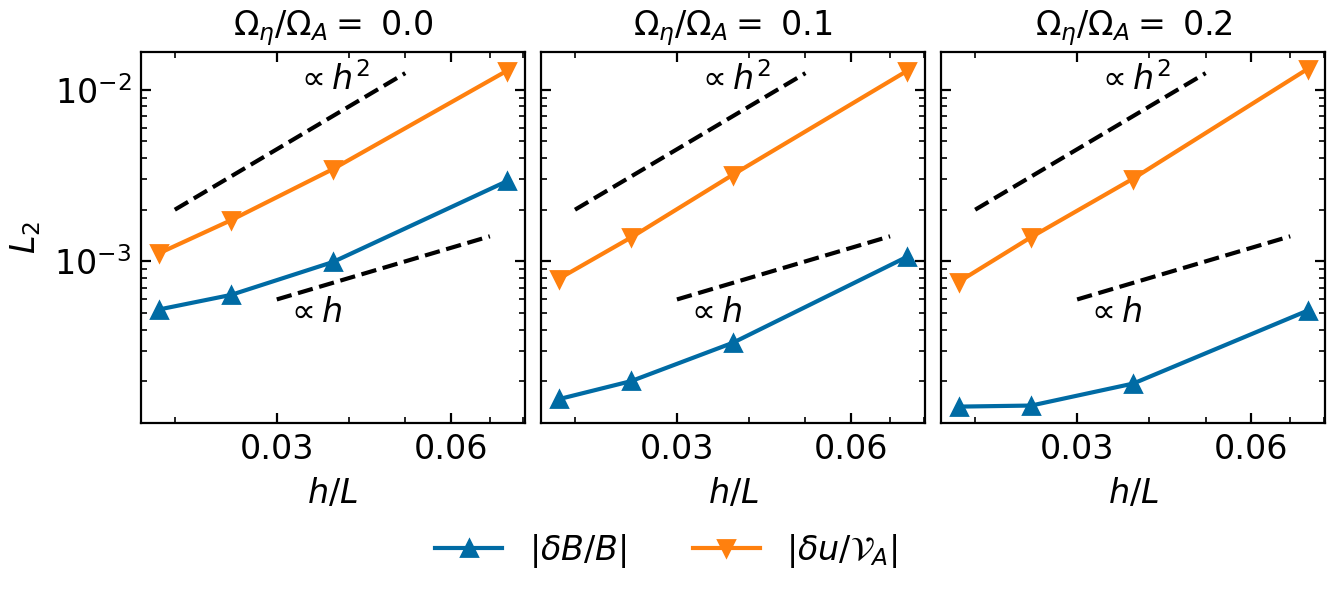}
 \caption{Convergence with resolution, in terms of smoothing length $h$ scaled to the size of the simulation box $L$, of the error on the calculated $\delta B / B$ and $\delta u / \mathcal{V}_A$. We include lines scaling as $\propto h$ and $\propto h^2$, corresponding to first and second-order convergence behaviour, to guide the eye. The error on the velocity perturbation seems to converge close to quadratically with resolution. The error on the magnetic field perturbation seems to converge more slowly, plateauing at a low value for the highest spatial resolutions we considered.}
 \label{fig:convergence}
\end{figure*}
 
Figure~\ref{fig:results} shows good agreement between numerics and theory for all Ohmic diffusion strengths considered: this is a non-trivial feat as discrete (magneto)hydrodynamics methods are inherently diffusive and dispersive. For instance, \cite{2005MNRAS.364..384P}, and later \cite{2018PASA...35...31P} (although with largely improved results) show that even simple SPMHD simulations of circularly polarised, non-cosmological, non-diffusive Alfvén waves suffer from both amplitude and phase errors that accumulate as time integration proceeds; they demonstrate that these can already be observed after a few wave periods, but are reduced as spatial resolution is increased. We seem to be only minimally affected by such effects. We nevertheless seek to quantify this through a convergence study: we repeat our experiment by keeping all problem parameters fixed to the values given in Section~\ref{sec:initialisation}, except for the number of resolution elements along the $x$ direction, which we vary from $16$ to $64$ in powers of $2$\footnote{This is common practice in SPH convergence studies, where the number of particles is only increased along the direction of interest for a given problem, see e.g. the test set-ups presented in Section 5.6 of ~\citealt{2018PASA...35...31P}.}. The profile of the deviation between numerics and theory being uniform in space at all times for all runs considered, we deem it sufficient to quantify the error in our calculations through the difference between the calculated and expected wave amplitudes, which we subsequently integrate over all scale factors for which we have simulation outputs. We express the error on the scaled linear perturbation $\Delta_Q$ corresponding to physical attribute $Q$ (where $Q=[B_c, u]$) through an $L_2$ norm defined as
\begin{equation}
    L_2 \equiv \sqrt{\frac{1}{N_s} \sum_{j=1}^{N_s} {|\Delta_{Q, \mathrm{SWIFT}} (a_j) - \Delta_{Q,\mathrm{theory}} (a_j)|}^2}
\label{eq:l2_norm}    
\end{equation}
where the index $j$ runs over $N_s$ SWIFT snapshots, each output at a simulation time corresponding to scale factor $a_j$. 

We present the results of our convergence study in Figure~\ref{fig:convergence}, showing the deviation between theory and numerics for both the magnetic and velocity fields, for all Ohmic diffusion strengths considered. We express resolution, as is typical in the SPH literature, in terms of the smoothing length $h$, which sets the fundamental spatial scale used in the interpolation sums. We moreover include lines showing linear and quadratic scalings of the error with $h$ to guide the eye. Formal convergence is a non-trivial question to address in Lagrangian particle-based methods. SPH is expected to be, at best, second order accurate when calculating gradients using an optimal choice of discrete differential operators, and when the spatial distribution of particles under consideration is reasonably smooth and well behaved~\citep{1992ARA&A..30..543M}. The latter is admittedly our case but the former is unfortunately not, as we are bound to use a discrete Laplacian operator to calculate magnetic diffusion, which does not fulfil all requirements necessary to ensure second order convergence. Nevertheless, looking at Figure~\ref{fig:convergence}, we see that the $L_2$ norm on $\delta u / \mathcal{V}_A$ scales almost with $h^2$ for all $\eta$ considered. The $L_2$ norm on $\delta B_c / B_c$ displays convergence between first and second order, before plateauing at a low value at high resolutions, most notably at high Ohmic diffusion strengths.

\section{Conclusion}
\label{sec:conclusion}

Strong observational evidence has established that magnetic fields are present at all scales in the universe. While notoriously non-trivial to model, recent advances in the fields of discrete methods and computational algorithms have led to an increasing number of numerical astrophysical models beginning to account for their evolution and effect. We argue that this step-up in model complexity should be accompanied by thorough and extensive testing, to gain confidence in a given method's reliability and the accuracy of its predictions.

With accrued interest for cosmic magnetism stemming from foundational works such as~\cite{2010Sci...328...73N} (who provide evidence for the presence of strong magnetic fields in cosmic voids, favouring a seeding mechanism of cosmological origin), development of tools to solve the equations of cosmological MHD is vital to test theoretical predictions. In this work, we build on from~\cite{2022MNRAS.515.3492B} who note that simple, easily verifiable and tractable example problems for code performance evaluation of implementations of cosmological MHD are scarce in the literature. This leads them to derive a set of simple analytical expressions describing the evolution of Alfvén and magnetosonic waves in an expanding universe, and verify that the simulation code AREPO~\citep{2010MNRAS.401..791S} can reproduce them. We here extend their formalism to non-ideal MHD, deriving analytical expressions describing standing, linearly polarised Alfvén waves for a fluid where magnetic resistivity has not been neglected. We model these non-ideal effects through an Ohmic diffusion term with a spatially and temporally constant diffusion coefficient $\eta$, which we add to the time evolution equation of the magnetic field. We find that this leads to damping and dispersive effects on the waves, which we corroborate by running numerical experiments using a novel implementation of SPMHD in the astrophysical simulation code SWIFT. We find good agreement between simulations and theory, and show that the former quickly converge to the latter even for a modest number of resolution elements and therefore computational cost. \\

We suggest that the example we present in this work can be adopted by the community as part of standard cosmological MHD code testing practice. 

\section*{Acknowledgements}

The authors would particularly like to thank the referee, Thomas Berlok, for insightful and constructive comments which helped improve the quality and completeness of this manuscript. They would also like to acknowledge the fruitful discussions shared with Federico Stasyszyn, that took place during the development of SWIFT's SPMHD module and which made this work possible. They finally feel indebted to all other people in the SWIFT team who contributed with suggestions, comments and constructive exchanges on discrete methods and code development.

%%%%%%%%%%%%%%%%%%%%%%%%%%%%%%%%%%%%%%%%%%%%%%%%%%
\section*{Data Availability}

SWIFT is fully open-source; extensive documentation describing how to use it and details on how to download it can be found at~\url{https://www.swiftsim.com}. Both the MHD module and the diffusive cosmological Alfvén wave example will soon be made publicly available, in a forthcoming code version release. They can be accessed earlier upon request to the authors.



\bibliographystyle{mnras}
\bibliography{cosmoDiffusiveAlfvenWaves_revised}

%%%%%%%%%%%%%%%%%%%%%%%%%%%%%%%%%%%%%%%%%%%%%%%%%%

%%%%%%%%%%%%%%%%% APPENDICES %%%%%%%%%%%%%%%%%%%%%

\appendix

\section{Dissipative Alfvén waves in an EdS universe}
\label{app:alfven_waves_eds}

We present here the Alfvén wave perturbative solution to~(\ref{eq:comoving_mass_continuity})-(\ref{eq:comoving_energy}), assuming an EdS universe for which $\Omega_m=1$ (with all other cosmological parameters set to zero), leading to $\dot{a} = H_0 / \sqrt{a}$. Following exactly the same procedure as in Section~\ref{sec:alfven_waves}, we obtain the master equation for linear perturbations in the magnetic field
\begin{equation}
    \frac{\mathrm{d}^2}{\mathrm{d} a^2}
    \left( \frac{\delta B_c}{B_c} \right)
    + \left( \frac{3}{2a} + \frac{\Omega_\eta}{a^{3/2}} \right)
    \frac{\mathrm{d}}{\mathrm{d} a}
    \left( \frac{\delta B_c}{B_c} \right)
    + \frac{\Omega_A^2}{a^2} \frac{\delta B_c}{B_c} = 0
\label{eq:master_ode_a_eds}
\end{equation}
which under the coordinate transformation $\zeta = 2 \Omega_\eta / \sqrt{a}$, $\Delta_B = \zeta^r \delta B_c/ B_c$ for some $r\in \mathbb{C}$ becomes
\begin{equation}
    \zeta \frac{\mathrm{d}^2 \Delta_B} {\mathrm{d} \zeta^2}
    - \left( 2r + \zeta \right)
    \frac{\mathrm{d} \Delta_B}{\mathrm{d} \zeta}
    + \bigg\{ r + \frac{1}{\zeta} 
    ( r^2 + r + 4 \Omega_A^2 ) 
    \bigg\} \Delta_B
    = 0.
\label{eq:master_ode_zeta_eds}
\end{equation}
For
\begin{equation}
    r = - \frac{1}{2} \pm 2 i \kappa
    \quad \text{where} \quad
    \kappa = \sqrt{\Omega_A^2 - \frac{1}{16}}
\end{equation}
(\ref{eq:master_ode_zeta_eds}) reduces to Kummer's equation \citep[see again][]{1972hmfw.book.....A}:
\begin{equation}
    z \frac{\mathrm{d}^2 w} {\mathrm{d} z^2}
    + (b - z) \frac{\mathrm{d} w} {\mathrm{d} z}
    - a w = 0
\end{equation}
for $a, b \in \mathbb{C}$, which has as a solution the confluent hypergeometric function
\begin{equation}
    {}_1 F_1 (a ; b ; z) = \sum_{n=0}^{\infty}
    \frac{a^{(n)} z^n}{b^{(n)} n!}
\end{equation}
with the Pochhammer function $a^{(n)}$ defined through
\begin{align}
\begin{split}
    a^{(0)} & = 1 \\
    a^{(n)} & = a (a+1) (a+2) \dots (a+n-1)
\end{split}
\end{align}
Putting all of the above together, we can formulate the solution to~(\ref{eq:master_ode_a_eds}) as
\begin{align}
\begin{split}
    \frac{\delta B_c}{B_c} (a) = 
    a^{-1/4} \bigg\{
    & c_1 e^{i \kappa \mathrm{ln} (a)}
    {}_1 F_1 \left(
    \frac{1}{2} - 2i \kappa ; 1 - 4i \kappa ; \frac{2 \Omega_\eta}{a^{1/2}}
    \right)
    \\
    & + c_2 e^{ - i \kappa \mathrm{ln} (a)}
    {}_1 F_1 \left(
    \frac{1}{2} + 2i \kappa ; 1 + 4i \kappa ; \frac{2 \Omega_\eta}{a^{1/2}}
    \right)
    \bigg\}
\label{eq:alfven_wave_B_solution_eds}
\end{split}
\end{align}
where $c_1$ and $c_2$ are integration constants. Expression~(\ref{eq:alfven_wave_B_solution_eds}) is the same as the solution presented by \cite{2022MNRAS.515.3492B}, with the only difference that here the two independent solution terms are further modulated by confluent hypergeometric functions dependant on $\Omega_\eta$. We further note that our solution reduces to theirs in the limit $\eta \rightarrow 0$, since ${}_1 F_1 (a ; b ; z) \rightarrow 1$ as $z \rightarrow 0$. Finally, it is also worth mentioning that the weights owing to magnetic diffusion here fall under the special case of
\begin{equation}
    {}_1 F_1 (a ; 2 a ; z)
    = e^{x/2} 
    {\left( \frac{x}{4} \right)}^{1/2 - \alpha}
    \Gamma \big( \alpha + \frac{1}{2} \big)
    I_{a - 1/2} \left( \frac{x}{2} \right)
\end{equation}
for $\Gamma(\cdot)$ the gamma function and $I_{\alpha}$ the modified Bessel function of the first kind of order $\alpha$. Exploring this further is beyond the scope of the current paper.

\section{Dissipative Alfvén waves in an EdS universe, with a scale factor dependent diffusion coefficient}
\label{app:alfven_waves_eds_bis}

As mentioned in the main text, a possible workaround to obtain simple, closed-form Alfvén wave perturbative solutions to~(\ref{eq:comoving_mass_continuity})-(\ref{eq:comoving_energy}) when assuming an EdS universe (for which $\Omega_m=1$ and all other cosmological parameters are set to zero, leading to $\dot{a} = H_0 / \sqrt{a}$), would be to relax the assumption that $\eta$ be scale factor independent. We can define a co-moving counterpart $\eta_c$ to $\eta$ as
\begin{equation}
    \eta_c = a^{-1/2} \eta
\label{eq:comoving_eta}
\end{equation}
given which, after following the exact same procedure as in Section~\ref{sec:alfven_waves}, we obtain the master equation for linear perturbations in the magnetic field
\begin{equation}
    \frac{\mathrm{d}^2}{\mathrm{d} a^2}
    \left( \frac{\delta B_c}{B_c} \right)
    + \left( \frac{3}{2} + \Omega_{\eta,c} \right) \frac{1}{a}
    \frac{\mathrm{d}}{\mathrm{d} a}
    \left( \frac{\delta B_c}{B_c} \right)
    + \left( \Omega_A^2 + \frac{\Omega_{\eta,c}}{2} \right) \frac{1}{a^2}
    \frac{\delta B_c}{B_c} = 0
\label{eq:master_ode_a_eds_comoving_eta}
\end{equation}
where
\begin{equation}
    \Omega_{\eta,c} 
    \equiv \frac{k \mathcal{V}_{\eta,c}}{H_0}
    \equiv \frac{k^2 \eta_c}{H_0}
\label{eq:comoving_omega_eta}
\end{equation}
(\ref{eq:master_ode_a_eds_comoving_eta}) is an Euler differential equation with a known, closed-form solution. It moreover is of the same form as the master ordinary differential equations leading to Alfvén (for the case of ideal MHD, optionally coupled to Navier-Stokes viscosity) and magnetosonic (for the case of ideal MHD with $\gamma=4/3$) wave solutions derived in~\cite{2022MNRAS.515.3492B}. We follow standard procedure as described in e.g.~\cite{Asmar2016-hm} and find the solution to~\ref{eq:master_ode_a_eds_comoving_eta} to be
\begin{equation}
    \frac{\delta B_c}{B_c} (a) =
    a^{-\Gamma_\eta^+} 
    \left( c_1 e^{i \kappa_\eta \mathrm{ln} a}
    + c_2 e^{- i \kappa_\eta \mathrm{ln} a}
    \right)
\label{eq:alfven_wave_B_solution_eds_comoving_eta}
\end{equation}
where $c_1, c_2$ are integration constants and where we have defined the diffusion-dependent factors
\begin{equation}
    \Gamma_\eta^\pm = \frac{1}{4}
    \left( 1 \pm 2 \Omega_{\eta,c}
    \right)
\end{equation}
and
\begin{equation}
    \kappa_\eta = 
    \sqrt{\Omega_A^2 - {\Gamma_\eta^-}^2}
\end{equation}
The solution for the velocity perturbation can subsequently be obtained through differentiation: plugging~(\ref{eq:alfven_wave_B_solution_eds_comoving_eta}) into~(\ref{eq:perturbed_induction}) one finds
\begin{equation}
\begin{split}
    \frac{\delta w}{\mathcal{V}_A} (a)
    = \frac{i}{\Omega_A} a^{-\Gamma_\eta^+}
    \Biggl[
    & c_1 \left( \Gamma_\eta^- - i \kappa_\eta
    \right)
    e^{i \kappa_\eta \mathrm{ln} a} & \\
    & + c_2 \left( \Gamma_\eta^- + i \kappa_\eta
    \right)
    e^{- i \kappa_\eta \mathrm{ln} a}
    \Biggr]
\label{eq:alfven_wave_u_solution_eds_comoving_eta}
\end{split}    
\end{equation}
Given~(\ref{eq:alfven_wave_B_solution_eds_comoving_eta}) and~(\ref{eq:alfven_wave_u_solution_eds_comoving_eta}), one can again find the integration constants expressed in terms of the initial perturbation amplitudes $A_B$ and $A_u$ by solving a couple system of linear equations with two unknowns. Going though with this exercise gives us
\begin{equation}
    c_1 = \frac{a_i \Gamma_\eta^+}{2 \kappa_\eta}
    e^{- i \kappa_\eta \mathrm{ln} a_i}
    \left[
        A_u \Omega_A \sqrt{a_i} + 
        A_B \left( \kappa_\eta - i \Gamma_\eta^- \right)
    \right]
\end{equation}
and 
\begin{equation}
    c_2 = \frac{a_i \Gamma_\eta^+}{2 \kappa_\eta}
    e^{i \kappa_\eta \mathrm{ln} a_i}
    \left[
        - A_u \Omega_A \sqrt{a_i} + 
        A_B \left( \kappa_\eta + i \Gamma_\eta^- \right)
    \right]
\end{equation}
which in turn leads to the solutions
\begin{equation}
\begin{split}
    \frac{\delta B_c}{B_c} (a) =
    \Biggl( \frac{a}{a_i} \Biggr)^{-\Gamma_\eta^+}
    \Biggl\{ 
    & A_B \mathrm{cos} (\psi) \\
    & +
    \frac{1}{\kappa_\eta}
    \Bigl(
    A_B \Gamma_\eta^- +
    i A_u \Omega_A \sqrt{a_i} \Bigr)
    \mathrm{sin} (\psi)
    \Biggr\}
\end{split}
\label{eq:alfven_wave_B_solution_eds_comoving_eta_icd}
\end{equation}
and
\begin{equation}
\begin{split}
    \frac{\delta u}{\mathcal{V}_A} (a) =
    \Biggl( \frac{a}{a_i} \Biggr)^{-\frac{1}{2} - \Gamma_\eta^+}
    \Biggl\{ 
    & A_u \mathrm{cos} (\psi) \\
    & -
    \frac{1}{\kappa_\eta}
    \Bigl(
    A_u \Gamma_\eta^- -
    \frac{i A_B \Omega_A}{\sqrt{a_i}}
    \Bigr) \mathrm{sin} (\psi)
    \Biggr\}
\label{eq:alfven_wave_u_solution_eds_comoving_eta_icd}
\end{split}
\end{equation}
where we have defined
\begin{equation}
    \psi = \kappa_\eta \mathrm{ln} (a / a_i)
\end{equation}
One can calculate the associated wave frequency by taking the time derivative of the phase $\psi$ to find
\begin{equation}
    \dot{\psi} = k \mathcal{V}_A a^{-3/2} 
    \sqrt{1 - \left( \frac{\Gamma_\eta^-}{\Omega_A}\right)^2}
\end{equation}
which is in line with standard MHD theory, up to the factor under the square root which arises because of us having accounted for cosmic expansion. This factor leads to $\dot{\psi} / k$ explicitly depending on $k$, which translates to the waves being dispersive~\citep[for a more elaborate discussion on this see again][]{2022MNRAS.515.3492B}. We finally note that in the ideal MHD limit
\begin{equation}
    \underset{\eta \rightarrow 0}{\mathrm{lim}}
    \Omega_{\eta, c} = 0
    \quad
    \text{and}
    \quad
    \underset{\eta \rightarrow 0}{\mathrm{lim}} \Gamma_\eta^\pm = \frac{1}{4}
\end{equation}
and~(\ref{eq:alfven_wave_B_solution_eds_comoving_eta_icd}) and~(\ref{eq:alfven_wave_u_solution_eds_comoving_eta_icd}) reduce to the corresponding solutions derived in~\cite{2022MNRAS.515.3492B}.
Choosing $A_B = 0$ leads to considerably simpler expressions that read
\begin{equation}
\begin{split}
    \frac{\delta B_c}{B_c} (a) =
    A_u
    \Biggl( \frac{a}{a_i} \Biggr)^{-\Gamma_\eta^+}
    &
    \frac{i \Omega_A \sqrt{a_i}}{\kappa_\eta}
     \mathrm{sin} (\psi)
\end{split}
\label{eq:alfven_wave_B_solution_eds_comoving_eta_icd_simple}
\end{equation}
and
\begin{equation}
\begin{split}
    \frac{\delta u}{\mathcal{V}_A} (a) =
    A_u
    \Biggl( \frac{a}{a_i} \Biggr)^{-\frac{1}{2} - \Gamma_\eta^+}
    \Biggl\{ 
    \mathrm{cos} (\psi) -
    \frac{\Gamma_\eta^-}{\kappa_\eta}
    \mathrm{sin} (\psi)
    \Biggr\}
\end{split}
\label{eq:alfven_wave_u_solution_eds_comoving_eta_icd_simple}
\end{equation}

Equations~(\ref{eq:alfven_wave_B_solution_eds_comoving_eta_icd_simple}) and~(\ref{eq:alfven_wave_u_solution_eds_comoving_eta_icd_simple}) can straightforwardly be used for a quantitative assesment of SWIFT results for simulations of standing, linearly polarised, cosmological Alfvén waves in an EdS universe. To initialise these we follow the same procedure as for the case of radiation-domination, as detailed in Section~\ref{sec:initialisation}, adopting the exact same parameters with the only exception of $\Omega_a$ which we here set to $\Omega_A = \pi$. This is mostly motivated by a desire to maximise readability of the resulting plots while still capturing the sought after features, given the vastly different time evolution of the scale factor ensuing from our change of cosmology.

\begin{figure*}
 \includegraphics[]{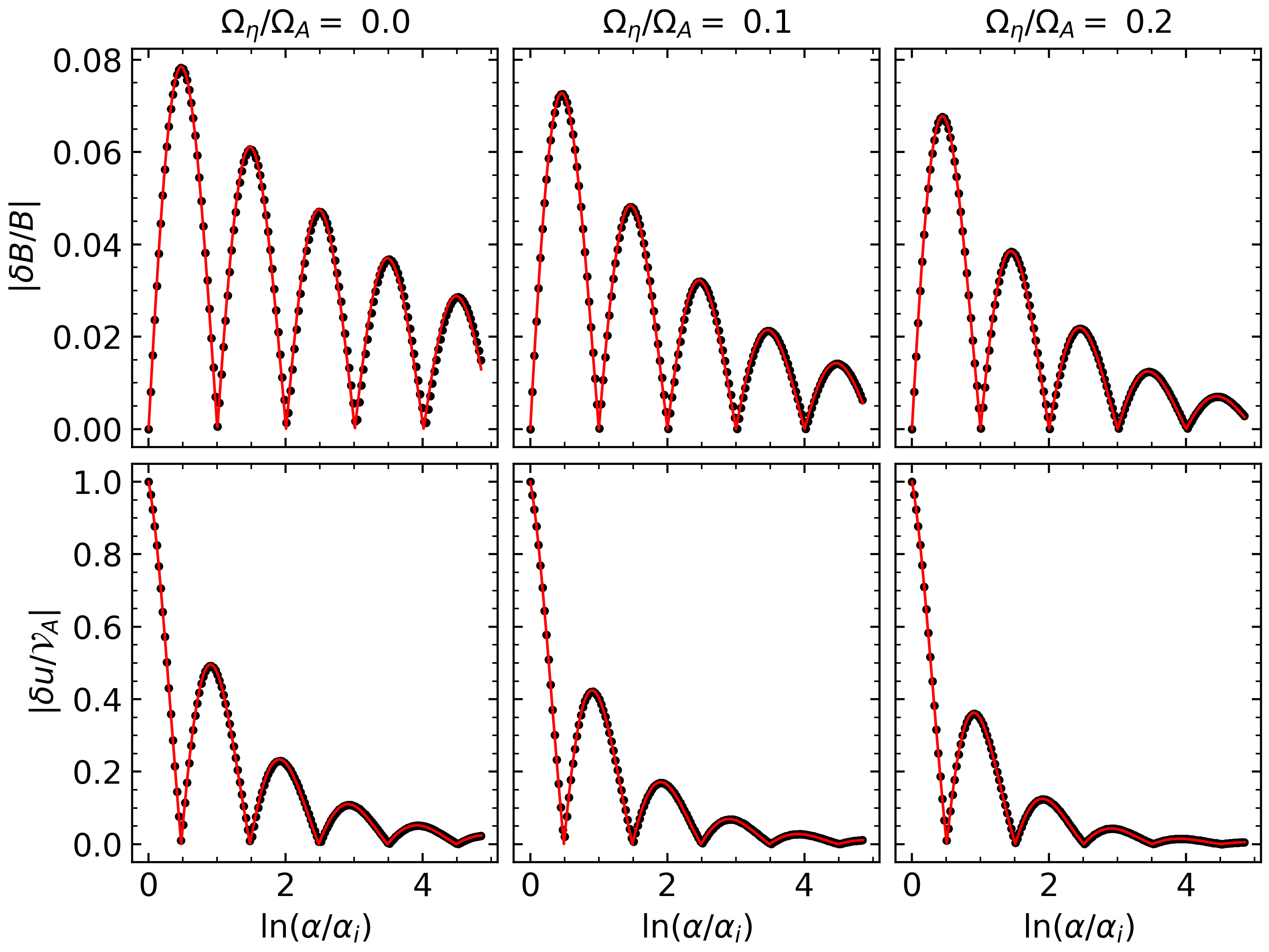}
 \caption{Analytical solutions (red line) plotted against numerical results obtained with SWIFT (black dots) for the time evolution in an EdS universe of the Fourier amplitude of magnetic (top row) and velocity (bottom row) linear perturbations corresponding to a standing, linearly polarised, cosmological Alfvén wave initialised as described in Section~\ref{sec:initialisation}. Individual columns correspond to different strengths of the Ohmic diffusion coefficient $\eta$ as expressed through the ratio $\Omega_\eta / \Omega_A$. The numerical solution agrees well with the analytically derived result, showing an increasingly strong damping (reduction of the wave amplitude) as $\eta$ is increased. The dispersive behaviour observed in the radiation dominated case is still present, but considerably less pronounced. We note that here, as opposed to Fig.~\ref{fig:results}, we choose as our abscissa $\mathrm{ln} (\alpha / \alpha_i)$ rather than $\alpha$.}
 \label{fig:results_EdS}
\end{figure*}

We compare the analytical solution we derived to numerical results obtained with SWIFT in Fig.~\ref{fig:results_EdS}, where we show the evolution as a function of \textit{the natural logarithm of the scale factor} (a natural variable choice given the functional form of~(\ref{eq:alfven_wave_B_solution_eds_comoving_eta_icd_simple}) and~(\ref{eq:alfven_wave_u_solution_eds_comoving_eta_icd_simple})) and for different Ohmic diffusion strengths of $\delta B_c / B_c$ and $\delta u / \mathcal{V}_A$.
As in the case of a radiation-dominated uinverse, solutions here take the form of a (mildly) dispersive oscillation, modulated by a decaying envelope the form of which depends on the value of $\eta$, for both variables considered. Here too SWIFT results agree well with theoretical expectations.

\begin{figure*}
 \includegraphics[]{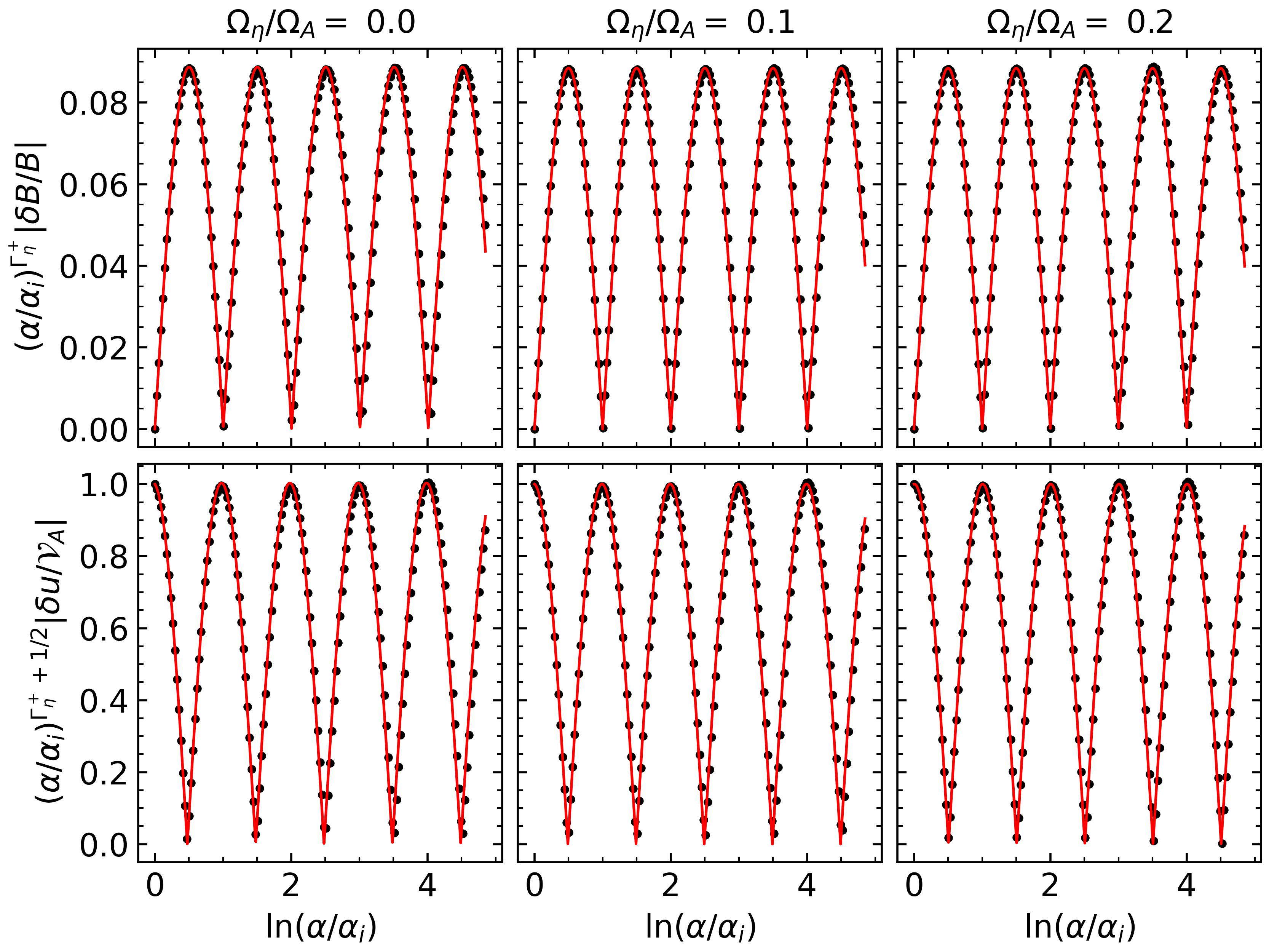}
 \caption{Same as Fig.~\ref{fig:results_EdS}, except that we have rescaled the magnetic and velocity perturbations by the cosmological and Ohmic diffusion damping terms appearing as multiplicative pre-factors in equations~(\ref{eq:alfven_wave_B_solution_eds_comoving_eta_icd_simple}) and~(\ref{eq:alfven_wave_u_solution_eds_comoving_eta_icd_simple}). Having factored out dissipation, our numerical results still follow the derived solution closely, for all Ohmic diffusion coefficients we consider.}
\label{fig:results_EdS_rescaled}
\end{figure*}

One advantage of the solutions derived here, as opposed to those presented in Section~\ref{sec:analytic_solutions}, is that they can be expressed in terms of elementary functions and, most importantly, the decay brought about by both cosmology and Ohmic diffusion is fully encoded as a single multiplicative pre-factor. It is then straightforward to divide it out of the solutions, to only be left with linear combinations of sines. We show a theory to numerics comparison for perturbations thus rescaled in Fig.~\ref{fig:results_EdS_rescaled}, to find once again good agreement between the two. As $\eta$ is varied, results presented this way appear indistinguishable.

\begin{figure*}
 \includegraphics[]{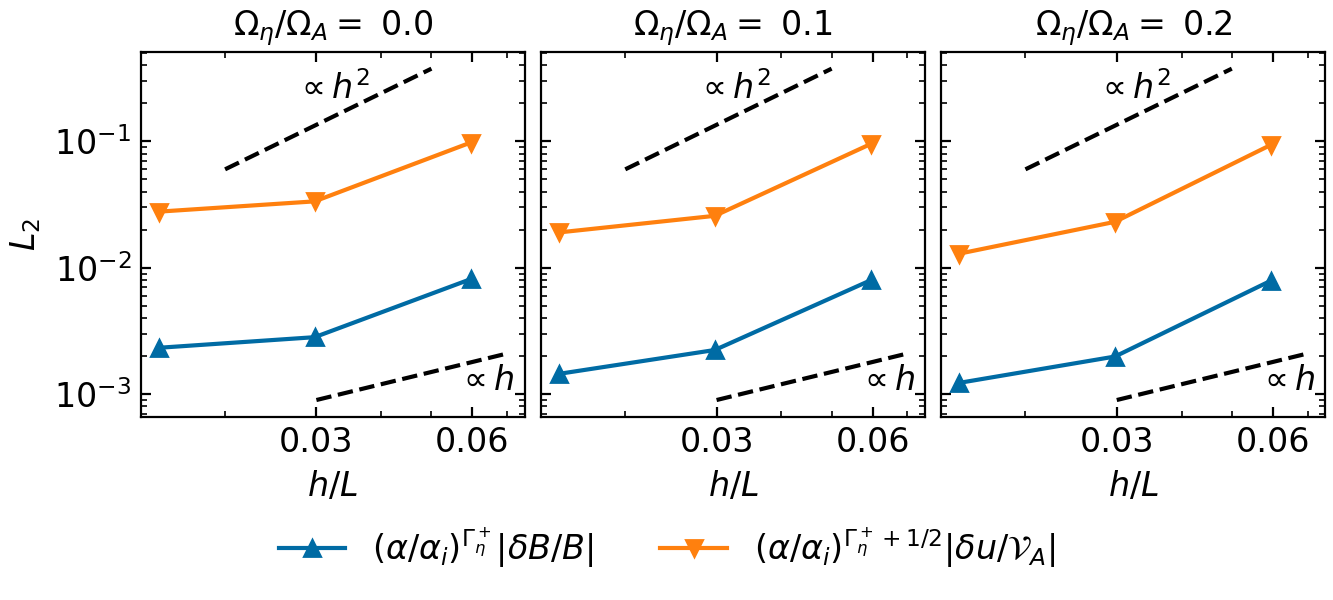}
 \caption{Convergence with resolution, in terms of smoothing length $h$ scaled to the size of the simulation box $L$, of the error on the calculated $\delta B / B$ and $\delta u / \mathcal{V}_A$ for which the decay due to cosmology and Ohmic diffusion has been factored out. We include lines scaling as $\propto h$ and $\propto h^2$, corresponding to first and second-order convergence behaviour, to guide the eye. Convergence seems to initially be close to quadratic, before eventually slowing down.}
\label{fig:convergence_EdS}
\end{figure*}

We perform a convergence study on the rescaled solutions for which, having factored out physical decay, we are only sensitive to numerical diffusion (which in SPH would solely be attributable to the corrective measures we implement, as the base algorithm is dissipationless) and to our method's systematics. We show our results in Fig.~\ref{fig:convergence_EdS} which, as also seen in Fig.~\ref{fig:results_EdS_rescaled}, seem largely insensitive to the choice of $\eta$. Similarly in a qualitative sense to what was observed in the radiation-domination case, the deviation between theory and numerics expressed in terms of an $L_2$ norm initially appears to scale close to quadratically with resolution before plateauing at a low value.

%%%%%%%%%%%%%%%%%%%%%%%%%%%%%%%%%%%%%%%%%%%%%%%%%%

% Don't change these lines
\bsp	% typesetting comment
\label{lastpage}
\end{document}